\documentclass[twocolumn,showpacs,amsmath,amssymb,pre,aps]{revtex4}   
\usepackage{graphicx}
\usepackage{dcolumn}
\usepackage{bm}
\usepackage{color}

\newif\ifgraph

\graphtrue             %

\begin{document}
\title{
Magnetic flux pinning in superconductors with 
hyperbolic-tesselation arrays of pinning sites
}
\draft

\author{V.~R.~Misko$^{1,2}$ and Franco Nori$^{1,3}$}
\affiliation{$^1$ Advanced Science Institute, RIKEN, Wako-shi, Saitama 351-0198, Japan}
\affiliation{$^{2}$Departement Fysica, Universiteit Antwerpen, 
B-2020 Antwerpen, Belgium}
\affiliation{$^3$ Physics Department, University of Michigan, 
Ann Arbor, MI 48109-1040, USA}

\date{\today}

\begin{abstract} 
We study magnetic flux interacting with arrays of pinning sites (APS) 
placed on vertices of hyperbolic tesselations (HT). 
We show that, due to the gradient in the density of pinning sites, 
HT APS are capable of trapping vortices for a broad range of applied magnetic fluxes. 
Thus, the penetration of magnetic field in HT APS is essentially different from the usual scenario predicted by the 
Bean model. 
We demonstrate that, due to the enhanced asymmetry of the 
surface barrier for vortex entry and exit, this HT APS 
could be used as a ``capacitor'' to store magnetic flux. 
\end{abstract}
 \pacs{
74.25.Wx 
02.40.-k; 
74.25.Uv; 
74.78.Na; 
}
\maketitle

\section{Introduction} 

Non-Euclidean geometries have had a very profound effect 
in physics~\cite{frankel}; 
not only in general relativity, but also 
more recently in condensed matter physics~\cite{nelson,sadoc,bowick}. 
Some examples include order and defects 
in liquids and metallic glasses~\cite{nelsonprb83}, 
polytope models of glass~\cite{nelsonnpb84} 
in a curved icosahedral space, 
icosahedral bond-orientational-order in supercooled 
liquids~\cite{steinhardt}, etc. 
Recently, topological defects, i.e., disclinations, dislocations 
and vortices, were studied on rigid substrates of spatially-varying 
Gaussian curvature~\cite{turner}. 
Here we present a study on vortex pinning in 
hyperbolic two-dimensional (2D) tesselations. 

In mesoscopic superconductors, 
the geometry of a sample and/or its underlying pinning has 
a strong impact on its vortex pattern; 
for example, the appearance of vortex concentric 
``shells''~\cite{irina,irina07,vitaly1998} 
in mesoscopic disks, 
which can even merge into so-called giant vortices~\cite{kanda04} 
in very small disks. 
In symmetric 
polygons, e.g., triangles and squares, vortices tend to form patterns with the symmetry of the polygon boundary. 
Moreover, the sample symmetry can even lead to a spontaneous 
generation of antivortices 
to restore the broken symmetry for incommensurate 
magnetic flux~\cite{vav,vavwe}. 
Furthermore, vortex patterns can also be produced by 
various arrays of pinning sites (APS) (see, e.g., \cite{kagome}). 
Even for the simplest case of a square APS, a variety of patterns 
and phases were found~\cite{walter09}. 

Using APS incommensurate with vortex lattices results 
in an elastic deformation of the vortex lattice 
and thus in an increase in the elastic energy. 
However, pinning properties of a superconductor can be even 
improved by using incommensurate APS sites, 
as recently demonstrated 
(theoretically~\cite{wepenprl,wepenprb} 
and experimentally~\cite{koelle,silhanek,vvmpen09,wepenprb2010}) 
for quasiperiodic (QP) tiling APS. 
The important property of 
QP APS~\cite{wepenprl,wepenprb,koelle,silhanek,vvmpen09,wepenprb2010} 
is the existence 
of many built-in periods resulting in 
flux pinning for various flux densities. 
This unique property, in turn, opens the possibility for 
the design of fluxonics devices with enhanced pinning 
over a broad range of fields. 

In this paper, we investigate pinning properties of a 
superconductor with pinning sites placed on the vertices 
of a {\it hyperbolic tesselation} (HT). 
One important property which makes this system different 
from the family of QP tilings, is that a HT 
(i.e., its projection from hyperbolic to a 2D Euclidean space, 
e.g., in a Poincar\'{e} disk representation) is a finite-size 
system. 
This makes HT somewhat similar to mesoscopic symmetric 
experimental samples~\cite{irina,irina07,kanda04,vav}, 
although with additional internal structure and less rigid 
``boundaries''. 
We show that an HT APS exhibits an enhanced asymmetry of its 
surface barrier, for vortex entry and expulsion. 

Hyperbolic tesselations are obtained as a projection 
of a tiling of regular symmetric polygons in a hyperbolic space 
to a two-dimensional Euclidean space. 
(Similarly, QP tilings are obtained as a 2D projection of 
tilings which are periodic in a higher dimensional space.) 
As a result, a 2D HT is a set of topologically equivalent tilings 
of decreasing sizes (from the center and towards the edge) 
which makes this system attractive 
for pinning magnetic flux, as we demonstrate below. 
Each tesselation is represented by a Schl\"{a}fli symbol of the form 
$\{p,q\}$, which means that $q$ regular $p$-gons surround each vertex.  
There exists a hyperbolic tesselation $\{p,q\}$ for every $p$, $q$ 
such that 
$(p-2)(q-2)>4$~\cite{frankel,nelson,sadoc,nelsonprb83,nelsonnpb84}. 
Note that the ones based on a regular $\{p,q\}$ are the same as the 
dual ones based on a regular $\{q,p\}$, but shown in a different orientation (see Fig.~1). 
\begin{figure}[btp] 
\begin{center} 
\includegraphics*[width=6.5cm]{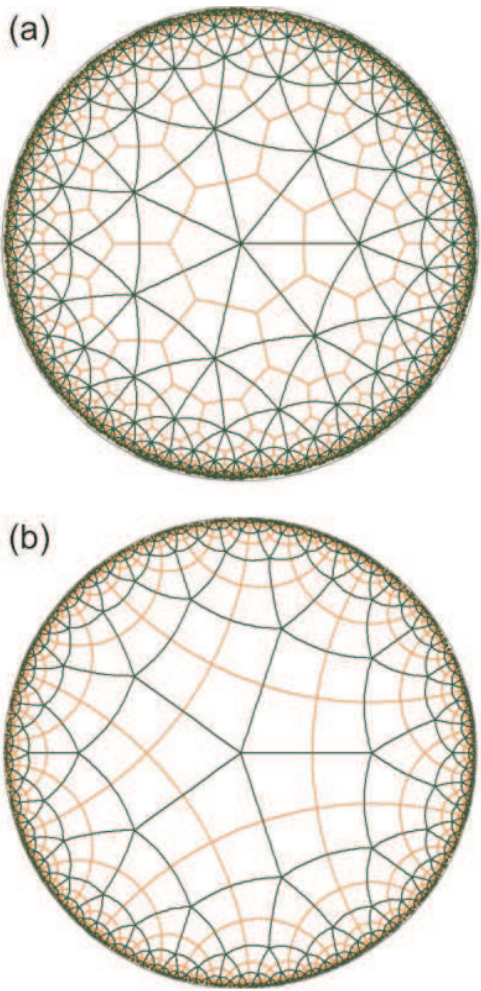} 
\end{center} 
\vspace{-0.5cm} 
\caption{ 
Hyperbolic tesselations in a Poincar\'{e} disk. 
The Shchl\"{a}fli symbols $\{p,q\}$ describe tesselations 
where $q$ regular $p$-gons meet at each vertex: 
$\{3,7\}$ (a) and $\{4,5\}$ (b) 
(shown by dark green lines). 
The corresponding dual tesselations 
$\{7,3\}$ (a) and $\{5,4\}$ (b) are 
shown by orange lines. 
} 
\end{figure}

\section{Simulation} 

We model a 3D column, infinitely long in the $z$-direction, 
by a 2D (in the $xy$-plane) square simulation cell (see Fig.~2) 
with periodic boundary conditions. 
The free-of-pinning region between the HT APS and the boundary 
of the simulation cell serves as a reservoir of vortices that 
mimics the externally applied magnetic field. 
This approach has been successfully used in numerous simulations 
in the past (see, e.g., 
\cite{FNprl94,FNsci97,FNprb97,wepenprl,wepenprb,wepenprb2010}). 
To study the dynamics of vortex motion, 
we numerically integrate the overdamped equations of motion 
(see, e.g., Refs.~\cite{wepenprl,wepenprb}): 
\begin{equation} 
\eta {\rm \bf v}_{i} \ = \ {\rm \bf f}_{i} \ = \ {\rm \bf f}_{i}^{vv} + {\rm \bf f}_{i}^{vp} + {\rm \bf f}_{i}^{T} + {\rm \bf f}_{i}^{d}.
\end{equation} 
Here
${\rm \bf f}_{i}$
is the total force per unit length acting on vortex
$i$,
${\rm \bf f}_{i}^{vv}$ 
and
${\rm \bf f}_{i}^{vp}$
are the forces due to vortex-vortex and vortex-pin interactions, respectively,
${\rm \bf f}_{i}^{T}$
is the thermal stochastic force,
and 
${\rm \bf f}_{i}^{d}$
is the driving force; 
${\rm \bf v}_{i}$
is the velocity, and 
$\eta$ is the viscosity. 
All the forces are expressed in units of
$
f_{0} = \Phi_{0}^{2} / 8 \pi^{2} \lambda^{3},
$
where $\Phi_{0} = hc/2e$, and
lengths (fields) 
in units of
$\lambda$ ($\Phi_{0}/\lambda^{2}$). 

Following 
Refs.~\onlinecite{FNprl94,FNsci97,FNprb97,wepenprl,wepenprb,wepenprb2010}, 
the force due to the interaction of the $i$-th vortex with other vortices is 
\begin{equation}
{\rm \bf f}_{i}^{vv} \ = \ \sum\limits_{j}^{N_{v}} \ f_{0} \ K_{1} \!
\left( \frac{ \mid {\rm \bf r}_{i} - {\rm \bf r}_{j} \mid }{\lambda} \right)
\hat{\rm \bf r}_{ij} \; , 
\label{fvv}
\end{equation}
where 
$N_{v}$ 
is the number of vortices, 
and 
$K_{1}$
is a first-order modified Bessel function. 

Vortex pinning is modeled by short-range parabolic potential wells 
located at positions 
${\rm \bf r}_{k}^{(p)}$. 
The pinning force is 
\begin{equation}
{\rm \bf f}_{i}^{vp} = \sum\limits_{k}^{N_{p}} \left( \frac{f_{p}}{r_{p}} \right)
\mid {\rm \bf r}_{i} - {\rm \bf r}_{k}^{(p)} \mid
\Theta \!
\left( 
\frac{r_{p} - \mid {\rm \bf r}_{i} - {\rm \bf r}_{k}^{(p)} \mid}{\lambda} 
\right)
\hat{\rm \bf r}_{ik}^{(p)},
\label{fvp}
\end{equation}
where 
$N_{p}$
is the number of pinning sites,
$f_{p}$
is the maximum pinning force of each potential well, 
$r_{p}$
is the range of the pinning potential, 
$\Theta$ 
is the Heaviside step function, 
and 
$\hat{\rm \bf r}_{ik}^{(p)} = ( {\rm \bf r}_{i} - {\rm \bf r}_{k}^{(p)} )
/ \mid {\rm \bf r}_{i} - {\rm \bf r}_{k}^{(p)} \mid.$

The temperature contribution ${\rm \bf f}_{i}^{T}$
is represented by a stochastic 
term obeying the following conditions: 
\begin{equation}
\langle f_{i}^{T}(t) \rangle = 0
\end{equation}
and
\begin{equation}
\langle f_{i}^{T}(t)f_{j}^{T}(t^{\prime}) \rangle = 2 \,  \eta \,  k_{B} \,  T \,  \delta_{ij} \,  \delta(t-t^{\prime}). 
\end{equation}

To obtain the ground state of a system of vortices, 
the system starts at 
some non-zero value of the ``temperature'' and gradually decrease 
it to zero, i.e., we perform a simulated-annealing simulation. 
This procedure mimics the annealing procedure in field-cooled experiments. 
This method was employed for the calculations of equilibrium 
vortex configurations to study pinning properties of the HT APS. 

For the simulation of flux entry, vortices were instead injected 
in the simulation region, at random positions at 
its boundary, thus mimicking an increasing external field. 
Correspondingly, flux expulsion was simulated by random removal 
of vortices from the region outside the HT, mimicking a 
decreasing external field. 
These simulations were performed at zero temperature.

\section{Vortex configurations in a Hyperbolic Tesselation APS} 

Let us consider vortices on a HT APS. 
One might naively expect that, 
in order to pin all the vortices, 
the ideal pinning potential should be a six-fold symmetric 
HT $\{3,6\}$. 
This tessellation is composed of triangles, such that six triangles 
join each vertex, like in usual Abrikosov lattices. 
However, configuration $\{3,6\}$ does not satisfy the condition 
$(p-2)(q-2) > 4$ and thus such HT does not exist. 

The so-called Omnitruncated tesselation $\{3,7\}$ is the 
``most nearly planar'' 
of all semiregular or regular hyperbolic tesselations, in the sense 
that if one tried to construct it from Euclidean planar polygons, 
the sum of the angles at each vertex would be as small as possible,  while exceeding 360 degrees. 
Therefore, we choose the $\{3,7\}$ HT for our numerical simulations 
(although other HT will possess similar pinning properties).

\begin{figure}[btp] 
\begin{center} 
\includegraphics*[width=5.5cm]{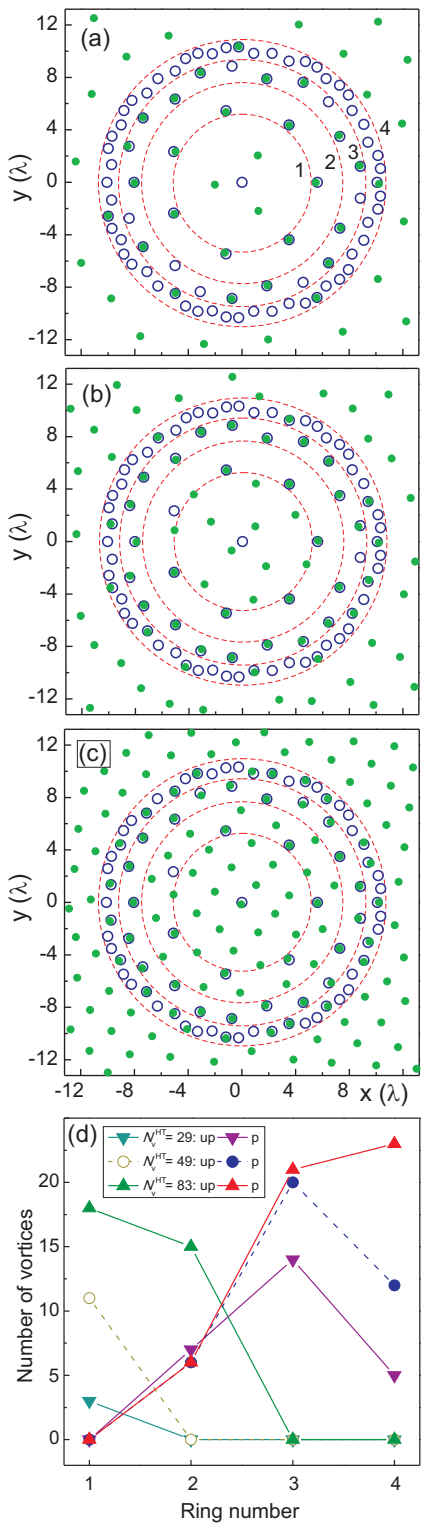} 
\end{center} 
\vspace{-0.5cm} 
\caption{ 
Vortex configurations in a $\{3,7\}$ hyperbolic tesselation (HT) 
APS
for varying number of vortices 
per simulation cell (inside the HT): 
$N_{v}=49$ ($N_{v}^{\rm HT}=29$) (a), 
$N_{v}=81$ ($N_{v}^{\rm HT}=49$) (b), and  
$N_{v}=144$ ($N_{v}^{\rm HT}=83$) (c). 
Pinning sites are shown by dark blue open circles, 
vortices by green dots. 
Red dashed lines show the boundaries of concentric 
rings [numbered 1 to 4 (a)] 
of the same area containing 1, 7, 21, and 56 
pins. 
(d) Number of pinned (p) and unpinned (up) vortices in the rings, 
corresponding to $N_{v}^{\rm HT}$ in (a to c), shown by symbols 
(lines connecting the symbols are guides for the eye). 
} 
\end{figure}

Figure~2 shows simulated stable vortex configurations 
in a $\{3,7\}$ HT APS 
for various values of the applied magnetic flux. 
As shown in Fig.~2(a), 
for low vortex densities 
(e.g., for $N_{v}=49$ vortices per simulation cell), 
vortices are mainly pinned by the central region of the HT. 
For higher densities (e.g., $N_{v}=81$), vortices are pinned by 
the central and intermediate region [Fig.~2(b)]. 
Finally, for even higher vortex density (e.g., $N_{v}=144$), 
the vortex lattice is commensurate with the pinning centers 
near the boundary, 
and therefore are more efficiently pinned in that region 
[Fig.~2(c)]. 

To analyze this observation in a more quantitative way, we divide 
the entire area of the HT in concentric rings of equal areas 
[Fig.~2(a)]. 
These rings of increasing radius contain 1, 7, 21, and 56 pinning 
sites, correspondingly (hereafter, we ignore the central 
pinning site assuming that the mobility of the vortices can be 
examined 
by rotating them with respect to the center of the HT, which can be 
experimentally realized in a Corbino setup --- see, e.g., 
Refs.~\cite{lin,okuma,linrat}). 
For the numbers of vortices considered here inside the HT 
(i.e., $N_{v}^{\rm HT}=29$, 49, and 83), 
the inner region contains 3, 11, and 18 unpinned vortices, 
correspondingly. 
The first inner ring accommodates, respectively, 7, 6, and 6 
pinned plus 15 unpinned vortices, i.e., all the vortices in 
case of low densities ($N_{v}^{\rm HT}=29$ and 49) in this central 
ring turn out to be pinned, while for the higher density, 
i.e., $N_{v}^{\rm HT}=83$, this ring is less efficient in terms 
of its pinning properties. 
The third ring, which is characterized by a higher density 
of pinning sites, is efficient for all the vortex densities 
considered, and it is able to trap more vortices with increasing 
vortex density. 
However, for the highest vortex density ($N_{v}^{\rm HT}=83$), 
most of the pinned vortices are pinned by the outer ring. 
These results are summarized in Fig.~2(d). 

The mobility of vortex matter pinned by the HT APS can be 
examined, e.g., by applying a Corbino-type radially 
decreasing external current 
(which, in addition, produces a shear stress) 
or simply by applying temperature. 
Thus, for low temperatures (and low vortex densities): 
the vortices at or near the center will be in a {\it liquid} 
state, as they are not pinned and can freely move; 
the vortices further away from the center will be in a 
{\it viscous liquid} vortex state, as they are only partially 
pinned (e.g., the vortices in the third ring for 
$N_{v}^{\rm HT}=83$); 
and the vortices located near the edges will be in a {\it solid} 
phase, because there are many pinning sites near the edges. 
Therefore, a single sample could exhibit three different vortex phases.

\section{Magnetic flux penetration in a HT APS} 

Due to the gradient in the density of pinning sites 
(which resulted, as shown above, in strongly inhomogeneous 
pinning of vortices), 
the penetration of magnetic flux in a HT APS is 
strongly inhomogeneous. 
The HT outer boundary has a high density of pinning sites 
(note that when approaching the boundary of a 
mathematical HT, the number of tiles and therefore the 
number of vertices goes to infinity, leading to an infinite 
density of pinning sites at the surface of the sample; 
however, here we consider a more realistic case 
of a finite number of vertices or pinning sites). 
The high-density-of-pins sample boundary 
shields the inner part of the HT APS by pinning vortices 
at the edge, thus preventing the magnetic flux from 
penetrating the HT. 

Figure~3 shows the penetration of magnetic flux into a HT APS 
for a number of vortices varying from 
$N_{v}=28$ [Fig.~3(a)] to 125 [Fig.~3(d)]. 
The initial state is prepared by annealing the vortices 
outside the HT such that they are homogeneously distributed 
in the region without pinning. 
After reaching a stable vortex configuration outside the 
``sample''
(which mimics a homogeneous externally-applied magnetic field), 
temperature is set to zero and 
vortices are allowed to freely move, e.g., 
enter the HT APS~\cite{FNprl94}. 
For low densities of the applied magnetic flux, 
i.e., ranging from $N_{v}=28$ to 63 [Figs.~3(a,b)], 
vortices are trapped by the dense pinning sites 
at the boundary. 
With increasing flux density, and thus flux ``pressure'', 
some vortices can pass over the first pinning row but still  
not able to enter the inner region of the HT, being pinned by 
the second row of pinning sites [Fig.~3(c) for $N_{v}=81$]. 
For higher flux densities, magnetic flux starts to slowly 
penetrate the HT APS via jumps from the outer 
pinning rows to the inner, less dense, pinning rows, 
while the edge pins trap additional incoming vortices, 
until all the pinning sites are filled. 
For even higher flux densities, vortices enter the central 
part of the sample where they self-organize in vortex 
patterns which are influenced by the circular boundary of 
the HT and the pinning sites inside the HT (in a similar 
way to the formation of vortex ``shells'' in mesoscopic 
superconducting disks with strong pinning~\cite{irina07}).

\begin{figure}[btp] 
\begin{center} 
\includegraphics*[width=6.5cm]{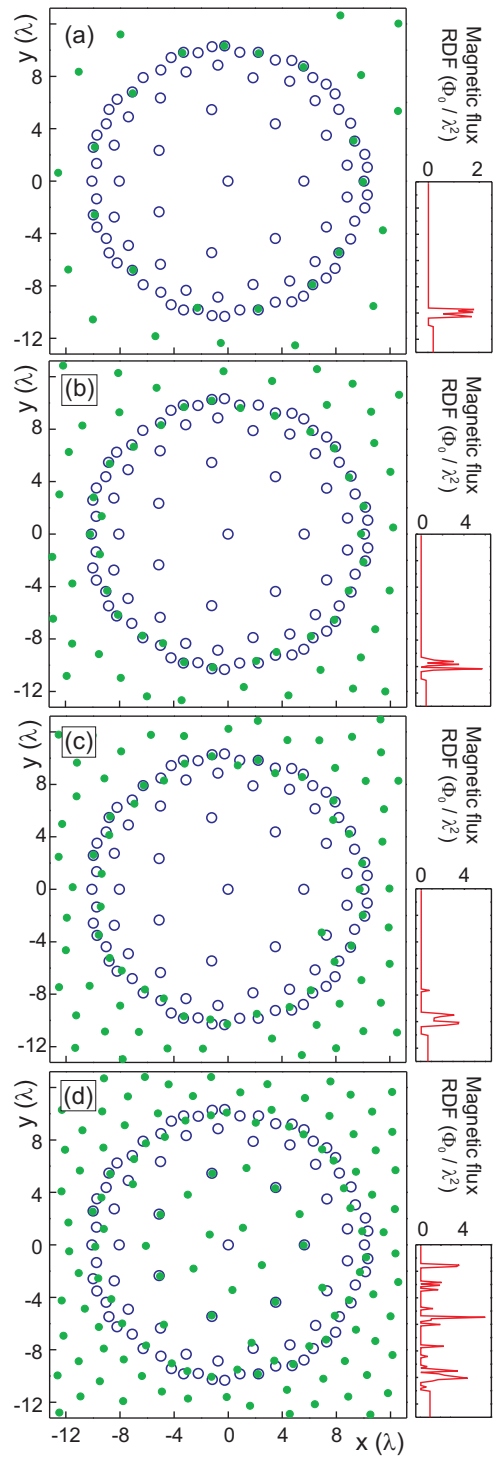} 
\end{center} 
\vspace{-0.5cm} 
\caption{ 
Penetration of magnetic flux in a HT APS: 
the ground-state vortex configurations for varying 
number of vortices,  
$N_{v}=28$ (a), 
63 (b), 
81 (c), and 
125 (d). 
For low and moderate applied fluxes (a, b), 
the penetration of the magnetic flux in the HT is shielded 
by the outer row (i.e., at the edge) of the HT, 
while for higher applied fluxes (c, d), 
vortices jump into the inner pinning sites, i.e., at the 
second pinning ``row'' and the pins at the central region, 
although vortices are still pinned by the outer pins. 
This scenario results in a very inhomogeneous magnetic flux 
penetration in the HT APS. 
The corresponding radial distribution function (RDF) 
of the magnetic flux penetrating in the HT APS 
is shown on the right-hand side panel for each $N_{v}$. 
} 
\end{figure}

The profile of the magnetic flux entering the HT APS 
as a function of the distance from the center of the HT, 
i.e., the radial distribution function (RDF), is shown 
in Figs.~3(e-h). 
The flux is accumulated in the vicinity of the sample 
boundary for moderate applied magnetic fluxes, and it 
penetrates deep inside the sample for larger fluxes. 
This scenario is in a sharp contrast to the conventional 
flux penetration described by the Bean model.

\section{Flux expulsion and accumulation} 

Another interesting property of the HT APS is the asymmetry 
of the sample boundary, i.e., inside and outside the HT. 
Of course, trapped flux in usual samples with no pinning 
is enhanced by the Bean-Livingston barrier~\cite{bl}. 
Adding pinning sites near the edges (as in a HT APS) 
strongly enhances this effect. 

To simulate flux expulsion and accumulation, 
the initial state is prepared by annealing the vortices 
{\it inside} the HT. 
After reaching a stable vortex configuration inside the HT, 
temperature is set to zero and vortices are allowed to move 
freely, i.e., they can leave the ``sample'' if the vortex density 
is sufficient. 
Magnetic flux accumulation in a HT APS and its expulsion 
is illustrated in Fig.~4.

\begin{figure}[btp] 
\begin{center} 
\hspace*{-0.5cm} 
\includegraphics*[width=7.0cm]{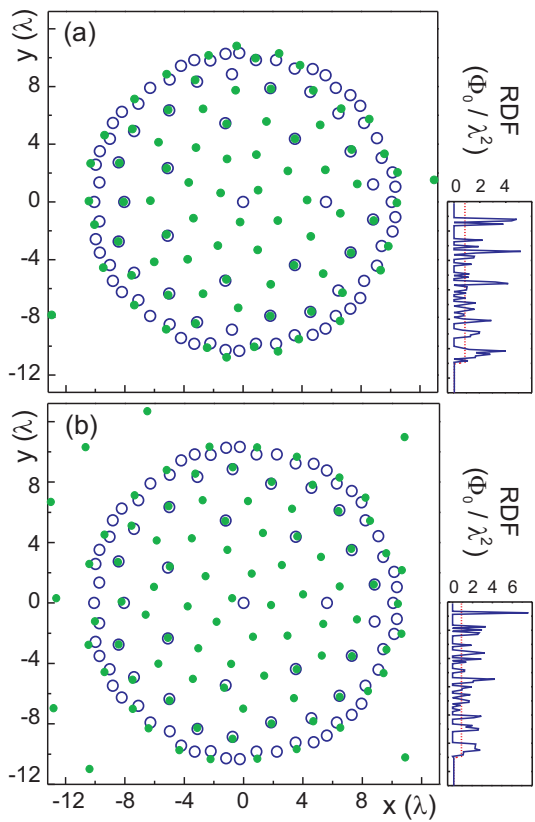} 
\end{center} 
\vspace{-0.5cm} 
\caption{
Accumulation of the magnetic flux in a HT APS 
and its expulsion: 
vortex patterns for 
$N_{v}=81$ (a), and 
93 (b). 
The right-hand side panels show the corresponding RDF for 
the magnetic flux inside ($r/\lambda \lesssim 11$) 
and outside ($r/\lambda \gtrsim 11$) the HT 
(solid dark blue line) and 
the average magnetic flux inside the HT 
(dashed red line). 
} 
\end{figure}

We analyzed the critical values for entry and expulsion 
of the magnetic flux and found that it is higher for the 
flux expulsion. 
For example, 
for an HT array of narrow pinning sites characterized by 
the maximum pinning force $f_{p}/f_{0}=2$ 
and radius $r_{p}/\lambda=0.3$, 
the ratio of the critical magnetic flux for flux expulsion 
to that of flux entry is $\Phi_{\rm out}/\Phi_{\rm in}\approx1.27$. 
This enhanced asymmetry in flux entry and expulsion might be useful 
for potential applications of HT APS as a ``capacitor'' type 
fluxonic device accumulating magnetic flux.

\begin{figure}[btp] 
\begin{center} 
\hspace*{-0.5cm} 
\includegraphics*[width=8.0cm]{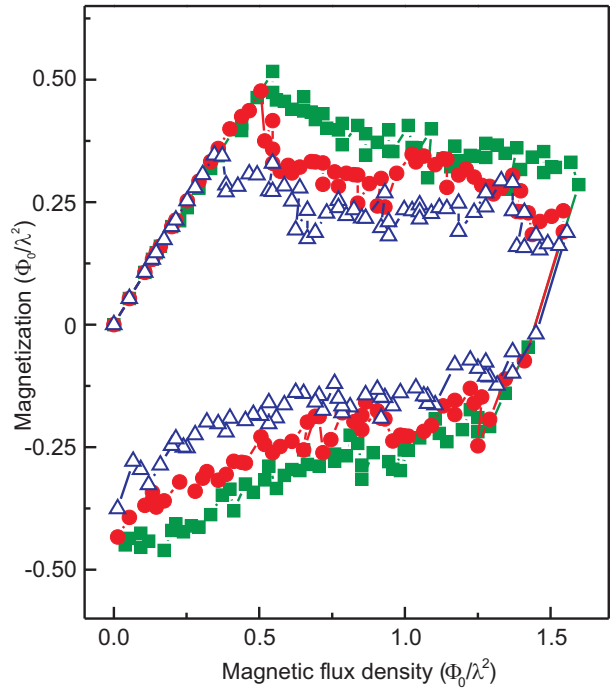} 
\end{center} 
\vspace{-0.5cm} 
\caption{ 
Magnetization curves 
for the HT APS 
with varying maximum pinning strength: 
$f_{p}/f_{0}=1.2$ (squares), 
1 (circles), and 
0.8 (open triangles). 
} 
\end{figure}

We calculated the magnetization of the sample 
exploiting the analogy between the vortices in the region 
outside the HT (i.e., free of pinning) and external field. 
The integrated difference between this field and the 
internal field~\cite{internal} 
across the sample yields the magnetization $M$~\cite{FNprl94}. 
In our calculations of the magnetization 
for the ramp-up phase, 
the number of vortices {\it outside} the HT was gradually increased 
from zero to $N_{v}^{\rm max}$, then it was gradually decreased 
to zero
for the ramp-down phase. 
Note that in this manner only a part of the magnetization loop 
can be calculated, i.e., for positive values of the magnetic 
field since the field cannot be inverted using this 
method~\cite{FNprl94}. 
Additional vortices were injected, one by one, at random positions 
at the boundaries of the simulation cell, providing a homogeneous distribution of the external magnetic field outside the sample. 

The calculations were performed for different pinning strengths. 
Typical magnetization curves are shown in Fig.~5. 
First, 
the magnetization 
$M$ increases linearly with the external field. 
It is interesting that the low-field slope of $M$ is equal to 1, 
reflecting the fact that magnetic flux does {\it not} penetrate 
the sample being shielded by the external boundary, 
i.e., the sample displays perfect diamagnetism. 
After reaching the maximum, $M$ starts to decrease 
due to vortices entering the sample. 
Note that this behavior of the magnetization 
in a HT APS is similar to 
that of a superconductor in a Meissner state. 
However, 
the magnetic flux is not ``expelled'' from the 
sample by a screening supercurrent as in usual 
superconductors in an external field. 
In a HT APS, the role of the screening current is played 
by the 
vortices which are pinned by the external row of 
pinning sites situated at the HT boundary. 
These pinned vortices repel incoming vortices and thus 
prevent the external flux from entering the sample. 
For various values of the pinning strength, 
the decrease in the magnetization, 
after reaching the maximum, 
can be either smooth 
(curves for $f_{p}/f_{0}=1.2$ and 0.8 in Fig.~5) 
or sharp 
($f_{p}/f_{0}=1$), 
depending on the parameters of the HT APS. 
Further exploiting the similarity to the magnetization 
of a superconductor in an external magnetic field, 
one can notice that 
this behavior resembles the magnetization in 
either 
{\it type-II} or 
{\it type-I} superconductors, 
although the reason for this behavior here can be due 
to commensurability effects. 

The magnetization of the HT has a remarkable hysteresis 
loop, as seen from the decreasing-field part of the 
magnetization curves in Fig.~5.

\section{Conclusions} 

We demonstrated that arrays of pinning sites (APS) placed 
on vertices of hyperbolic tilings, or tesselations (HT), 
can efficiently trap flux of different densities, in contrast 
to periodic APS which are efficient for few specific matching 
flux values. 
Vortex matter in this device can coexist in three different 
phases, namely, in a liquid phase (near the center), 
in a viscous liquid vortex phase (further from the center), 
and in a solid phase (near the boundary). 
This could be considered as a vortex analog of phases of matter 
inside some planets: 
a molten matter at the core, 
a viscous fluid surrounding the core, 
and a solid crust. 

We analyzed the magnetic flux entry and exit from a HT APS. 
For relatively low fluxes, the outer row of pinning sited 
shields the interior of the sample giving rise to 
a strongly inhomogeneous magnetic flux penetration, 
in contrast to that predicted by the conventional Bean model. 
The magnetization of this device has a linear part 
(for low applied magnetic fields) 
with the slope equal to one, which is indicative of 
a perfect diamagnetism similar to that in a Meissner 
state of a superconductor, although in a HT APS this 
occurs due to the shielding effect of trapped vortices 
in the outer pinning ring. 
Due to the asymmetry in flux entry and exit, 
the magnetization of the HT APS displays a remarkable hysteresis. 

Our predictions can be readily verified, e.g., in experiments 
on magnetization measurements. 
In particular, such measurements can be done using an array of HTs, 
to provide a sufficiently strong magnetization to be detected 
in experiments. 
Furthermore, our results for superconducting vortices can be 
easily extended to other systems of interacting particles, 
e.g., colloids in a HT potential landscape created by lasers. 
Our results are general since they do not depend 
on a specific form of interparticle interaction but 
rather reflect the more fundamental interplay between discrete 
periodic elastic media and incommensurate hyperbolic tilings.

\section{Acknowledgments} 
 
V.R.M. acknowledges support from 
the ``Odysseus'' Program of the Flemish Government 
\& FWO-Vl, and the IAP. 
F.N. is partially supported by the ARO, 
NSF grant No.~0726909, 
JSPS-RFBR contract No.~12-02-92100, 
Grant-in-Aid for Scientific Research (S), 
MEXT Kakenhi on Quantum Cybernetics, 
and the JSPS via its FIRST program.

\end{document}